\documentclass[letter]{IEEEtran}
\usepackage{ifpdf}
\usepackage{cite}
\ifCLASSINFOpdf
  \usepackage[pdftex]{graphicx}
  \DeclareGraphicsExtensions{.pdf,.jpeg,.png,.jpg}
\else
  \usepackage[dvips]{graphicx}
  \DeclareGraphicsExtensions{.eps}
\fi

\usepackage{graphicx}
\usepackage[cmex10]{amsmath}
\usepackage[tight,footnotesize]{subfigure}

\usepackage{stfloats}
\usepackage{url}
\usepackage[x11names,table]{xcolor}
\usepackage{multicol}
\usepackage{multirow}
\usepackage{amssymb}
\begin{document}
\pagestyle{empty}
\title{The Influence of the Buffer Size in Packet Loss for Competing Multimedia and Bursty Traffic}


%use this format for affiliations:

\author{\IEEEauthorblockN{Luis Sequeira, Juli\'{a}n Fern\'{a}ndez-Navajas, Luis Casadesus, Jose Saldana, Idelkys Quintana, Jos\'{e} Ruiz-Mas}
\IEEEauthorblockA{Communications Technology Group (GTC)-Arag\'{o}n Inst. of Engineering Research (I3A)\\
Dpt. IEC. Ada Byron Building. EINA. Univ. Zaragoza\\
50018 Zaragoza, Spain\\
Email: \{sequeira, navajas, jsaldana,jruiz,idelkysq,luis.casadesus\}@unizar.es\\}}


\maketitle
\thispagestyle{empty}

\begin{abstract}
%\boldmath
This work presents an analysis of the effect of the access router buffer size on packet loss rate and how it can affect the QoS of multimedia services when bursty traffic is present. VoIP traffic, real traces of viedoconferencing and videosurvellance are used in two different scenarios with medium link utilization. The study shows that the bursty nature of some applications may impair the MOS of voice calls especially when a certain number of bursts overlap. When link utilization is above $ 70\% $ good values of VoIP QoS cannot be obtained.

\end{abstract}


% \fi
\IEEEpeerreviewmaketitle

\section{Introduction}

A significant amount of network traffic \cite{games3, camera1} is originated nowadays by the new multimedia services over the Internet (e.g. videoconferencing, video surveillance, VoIP, online games and P2P-TV) and a large increase in the number of users can be observed. Moreover, the expectation of future growth by the use of multimedia applications, indicates that this tendency will increase in the next years. On the one hand, the user demands better experiences in multimedia services and on the other hand, the heterogeneous characteristics of the different Internet access technologies, make it necessary to take into account the Quality of Service (QoS) that they offer, especially when the accesses have to support real-time applications. 

At the same time, traffic behaviour may have a significant impact on network resources. The traffic behaviour provided by each service varies according to the nature of the information and its size, so different applications generate bursts of traffic (e.g. video surveillance) when a lot of information has to be sent in a short time. These bursts include different numbers of frames, and this may congest network devices when the amount of transmitted packets is significant with respect to the buffer size. On the other hand, some applications work to generate smooth traffic, with the aim of providing a certain QoS and a better user's experience, while not being detrimental for the network, at the cost of processing capacity increment. The size of the packets generated by these applications may vary between different Internet services: while some of them (e.g. VoIP) generate small packets of a few tens of bytes, others (e.g., videoconferencing and video surveillance) use large packets. By contrast, the buffer size and available bandwidth are maintained at the same value creating congestion problems on sensitive access links.

In this context, when we are planning a network, the size of the access router buffer is an important design parameter because there is a relationship between its size and link utilization, since when the buffer is full and the amount of memory is big, it would generate a significant latency increment (\textit{bufferbloat}). On the other hand, a very small amount of memory will increase packet loss in congestion time. As a result, the buffer behaviour is an important parameter which should be considered when trying to improve link utilization. In the last years, many studies related to buffer size issues have been published, but they are mainly focused on backbone routers and TCP flows \cite{buffers5}.

There are many techniques in order to improve link utilization but most of them are focused on bandwidth. Nevertheless bandwidth is not the only parameter to take into account. The buffer size and its behaviour are of primary importance when studying network traffic because buffer is used as a traffic regulator mechanism, since it may modify some network parameters, as delay or jitter, and may also drop packets. As a consequence, the influence of the buffer is important in order to reduce packet loss and offer a better user experience, especially when multimedia flows are being transmitted.

The paper is organized as follows: section II is a review of buffer dimensioning, section III discusses the influence of the buffer in different real-time services, to highlight the need of taking the buffer behaviour into account when planning a network. Section IV describes the specific congestion problems we are addressing in this paper and section V covers the experimental results. The paper ends with the conclusions.

\section{Buffer size issues}

\subsection{Buffer sizing}

Buffers are used to reduce packet loss by absorbing transient bursts of traffic when routers cannot forward them at that moment. This problem exists because the router has differences in the input and output rates that produces bottlenecks in the network, so packet loss may occur. Buffers are instrumental in keeping output links fully utilized during congestion times.

With respect to buffer dimensioning, the accepted \textit{rule of thumb} was using BDP (Bandwidth Delay Product) \cite{buffers2} as a method to obtain the buffer size needed at a router's output interface. This rule was proposed in $ 1994 $ \cite{buffers6} and it is given by $ B=C \times RTT $, where $ B $ is the buffer size, $ RTT $ is the average round-trip time and $ C $ the capacity of the router's network interface. It was experimentally obtained using at most $ 8 $ TCP flows on a $ 40 \; Mbps $ core link, so there is no recommendation for sizing buffers when there is a significant number of TCP flows with different $ RTTs $.

In \cite{buffers7}, the authors proposed a reduced buffer size by dividing BDP by the square root of the number of used TCP flows $ B=C \times RTT / \sqrt{N} $. This new approximation assumes that the number of TCP flows is large enough so as to consider them as asynchronous and independent from each other. This model was called \textit{small buffer} \cite{buffers8}.

In \cite{buffers10} it was suggested the use of even smaller buffers, called \textit{tiny buffers}, considering a size of some tens of packets. However, the use of this model presents a discarding packet provability of $ 10\%-20\% $. The model was obtained based in no bursty traffic. 

However, some real-time IP flows are bursty as e.g., video streaming, so this leaves some uncertainty in buffer sizing. In \cite{buffers9} and \cite{buffers5}, TCP and UDP combined traffic in very small buffers was tested using non-bursty traffic, and finding an anomalous region for UDP packets, where loss probability grows when buffer size increase.

It has also been observed in the literature that the buffer size is measured in different ways: e.g., in \cite{buffers1} the routers of two manufactures are compared, and one gives the information in packets, whereas the other one measures it in milliseconds, which is equivalent to bytes. As a consequence, the knowledge of the buffer behaviour is an interesting parameter which can be considered when trying to improve link utilization.

\subsection{Buffer overflow with medium link utilization}

\begin{figure}[t]
	\centering
	\includegraphics[height=4in,width=0.45\textwidth]{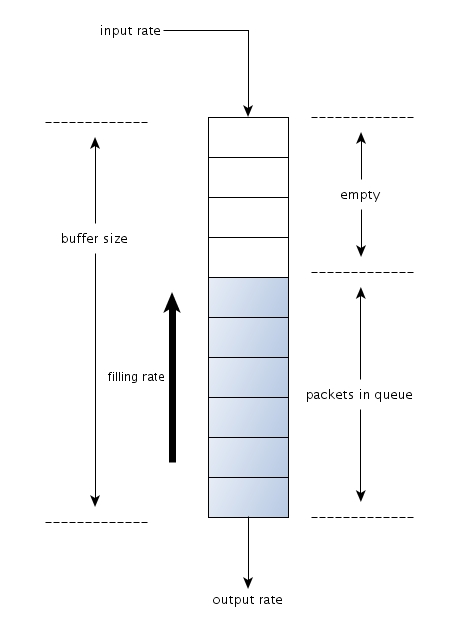}
	%\vspace{-0.35in}
	\caption{Principal characteristics of buffers.}
	\label{fig:buffer}
\end{figure}

The congestion of the buffer is not only caused by bandwidth scarcity but some problems can also be caused by network devices' implementations. As we can see in figure \ref{fig:buffer}, the time required to get into overflow is related to the filling rate, given by the input and output rate \cite{yo1}. Let $ R_{in} $ and $ R_{out} $ be the input and output rates of the buffer, respectively. We define $ R_{fill} $ as the rate in which the buffer fills when $ R_{in} $ is higher than $ R_{out} $ ($ R_{fill}=R_{in}-R_{out} $). 

So, when bursty traffic is generated in the network, a quick buffer filling rate may cause packet loss at certain moments, even when average link utilization is medium or even low. This may be caused when the burst length is nearly the buffer size because it gets easily into overflow, it also can be produced when burst length is bigger than buffer size, in this case packet loss will be automatic. 

It is known that some applications work to generate smooth traffic, but the aggregate Internet traffic shows a bursty behaviour at all time scales \cite{bursty1}. In this case, it is useful to manage traffic generation by taking advantage of traffic's smoothing configuration of certain applications. 

In some scenarios, when network congestion problems arise, a common practice can be increasing bandwidth in the local network. For this reason many companies change their internal network devices (e.g. switching from lowest to highest network rates) trying to resolve congestion problems. But, if $ R_{out} $ remains the same and $ R_{in} $ is switched to highest rate, buffer filling rate ($ R_{fill} $) is bigger in the new network. For this reason buffer may get into overflow more quickly. Thus, in certain cases this speed increase may be translated into a worse network response, so this improvement becomes a failure.

All in all, the relationship between Internet access and local network speeds and the relationship between buffer size and burst length are in fact an important parameters which cannot be neglected. In the next section, we present a number of tests with the aim of illustrating this phenomenon: some cases in which the combination of certain buffer sizes with bursty applications may cause network congestion and QoS problems.

\begin{figure*}[t]
	\centering
	\subfigure[Real videoconferencing traffic capture scenario.]{\includegraphics[height=2.5in,width=0.47\textwidth]{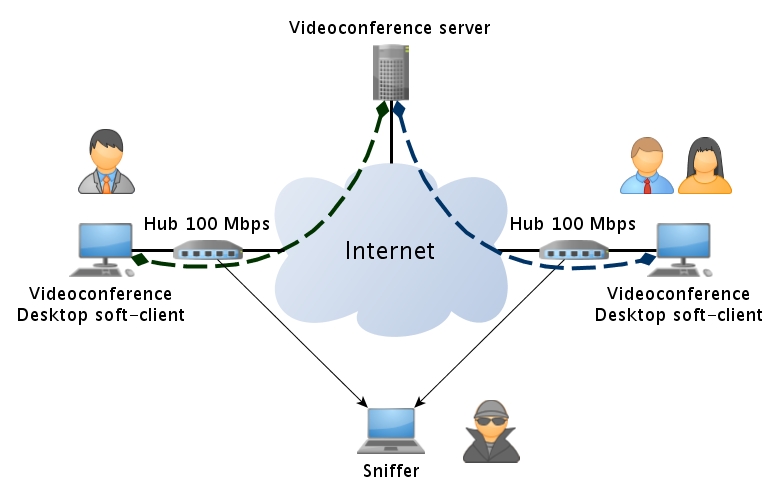}}
	\subfigure[Real video surveillance traffic capture scenario.]{\includegraphics[height=2.5in,width=0.47\textwidth]{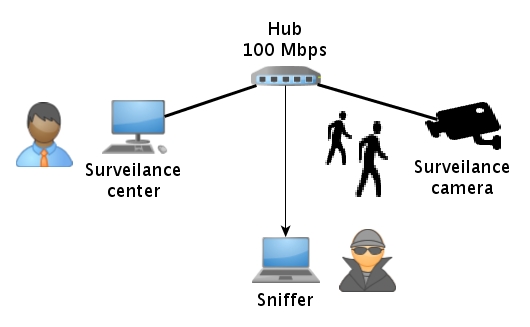}}
	%\vspace{-0.2in}
	\caption{Real traffic capture scenarios.}
	\label{fig:test_traffic}
\end{figure*}

\section{Review of the influence of buffers on different services}

Many scientific publications related to the influence of the buffer on different services and applications show how QoS is affected by the buffer behaviour, which is mainly determined by its size and management policies. In these cases, knowing the technical and functional characteristics of this device becomes important. This knowledge can be useful for applications and services in order to make decisions on the way the traffic is generated. In addition, packet management techniques can be applied as e.g. multiplexing a number of small packets into a big one or fragmentation, according to each case \cite{gtc17}.

The way to study the influence of the buffer for multimedia traffic is mainly to determinate QoS characteristics based on well known network parameters (e.g., jitter, packet loss, etc.) and also the subjective quality evaluation is used to determinate the users' perception for certain services. The ITU-T E-Model \cite{emodel} \cite{mos} has obtained a procedure with the aim of calculating the Mean Opinion Score, MOS, which is useful en network transmission planing. Others authors \cite{games5}, have developed a similar model for online games based on delay and jitter.

The influence of the buffer on VoIP was studied in \cite{gtc17}, where three different router buffer policies (dedicated, big and time-limited buffer) were tested, also using two multiplexing schemes. Router buffer policies cause different packet loss behaviour, and also modify voice quality, measured by means of R-factor. In the same paper a multiplexing method for VoIP flows is studied, thus reducing bandwidth with the counterpart of increasing packet size, which has an influence on packet loss, depending on the implementation and size of the router. In this case the VoIP native traffic showed a good behaviour when using a small buffer measured in bytes, as small packets have less probability of being discarded than big ones.

In \cite{gtc14} the authors present a simulation study of the influence of a multiplexing method on the parameters that define the subjective quality of online games, mainly delay, jitter and packet loss. The results show that small buffers present better characteristics for maintaining delay and jitter in adequate levels, at the cost of increasing packet loss. In addition, buffers whose size is measured in packets also increase packets loss.

Many access network devices are designed for bulk data transfers \cite{p2p4}, such as mail, web or FTP services. However, other applications (e.g., P2P video streaming, online games, etc.) generate a high rates of small packets, so the routers may experience problems to manage all the packets. In this case, their processing capacity can become a bottleneck if they have to manage too many packets per second \cite{games4}. The generation of hight rates of small packets \cite{p2p3y18} may penalize the video packets and consequently  peer's behaviour within P2P structure will not be as expected.

\section{Testbed and simulation results}

In this section, the phenomena derived from the buffer filling rate are studied in three test environments, using different buffer characteristics. These three scenarios have been simulated using NS-2 to analyze the effect of the buffer size in the presence of bursty traffic and the possible implications for the traffic of other applications. We will focus on a typical SME environment with one Internet access link. Different buffer sizes have been chosen according to suggestions of different works cited above and real sizes of commercial network devices (e.g., Linksys WAP54G) \cite{yo1}. 

\begin{figure*}[ht]
	\centering
	\subfigure[First scenario: two and three camera connections.]{\includegraphics[height=2in,width=0.47\textwidth]{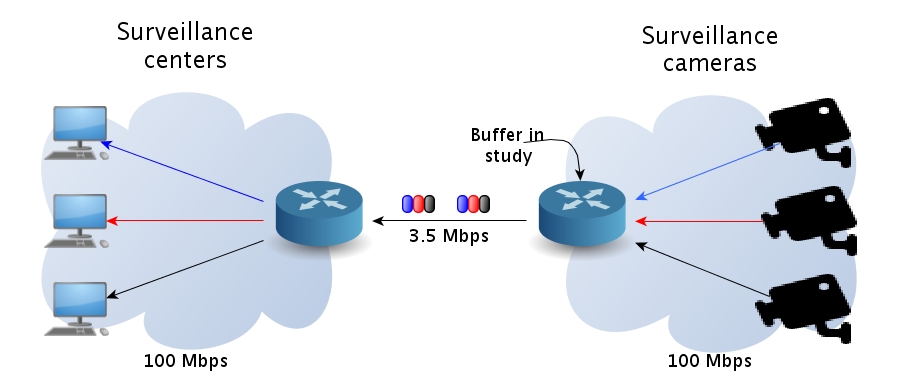}}
	\subfigure[Second scenario: two camera connections, videoconferencing and two VoIP calls.]{\includegraphics[height=2in,width=0.47\textwidth]{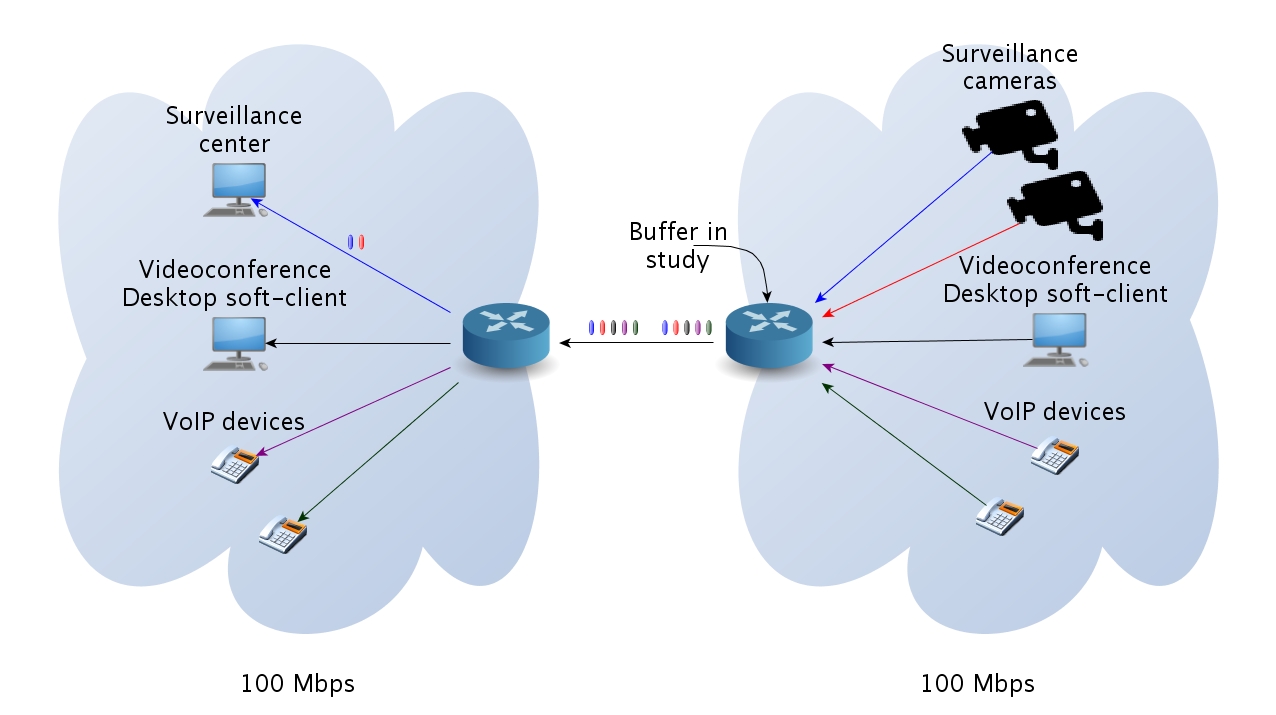}}
	%\vspace{-0.2in}
	\caption{Simulated scenarios.}
	\label{fig:scenarios}
\end{figure*}

\subsection{Traffic used}

We have used three different multimedia traffic sources: videoconferencing, video surveillance and VoIP. The methodology for obtaining the traffic captures can be seen in figure \ref{fig:test_traffic}. Real traces of videoconferencing and video surveillance applications were first captured in real scenarios and then generated in NS2, following the same packet sizes and inter-packet times. VoIP traffic has been generated with NS2 CBR agent. 

For videoconferencing traces, Vidyo\texttrademark architecture was used. Vidyo\texttrademark incorporates Adaptive Video Layering (AVL) technology which permits dynamical video optimization for each endpoint leveraging on $ H.264 $ Scalable Video Coding (SVC)-based compression technology. The videoconference software was configured with $ 2 \; Mbps $, $ 800 \times 450 $ resolution and the camera was capturing a high motion video (rugby game). 

%\begin{figure*}[ht]
%	\centering
%	\includegraphics[height=2.5in,width=0.45\textwidth]{Images/vidyo.jpg}
%	%\vspace{-0.35in}
%	\caption{Videoconferencing traffic capture scenario.}
%	\label{fig:vidyo}
%\end{figure*}

The video surveillance traffic traces have been obtained (see figure \ref{fig:test_traffic}) from a popular IP camera device (AXIS $ 2120 $). This kind of traffic is particularly bursty; table \ref{table:camera_packets} shows the relationship between the compression level and the amount of packets per burst for two different resolutions when bandwidth camera was set to $ 1 \; Mbps $. For all the test we have been chosen traces with $ 704 \times 576 $ resolution and a compression of $ 32 \; kbytes $, the time between bursts is $ 0.278 s \pm 0.06 s $, the amount of packets per burst is $ 26 $ and packet size is $ 1500 \; bytes $. 

%\begin{figure}[ht]
%	\centering
%	\includegraphics[height=2in,width=0.45\textwidth]{Images/test_traffic.jpg}
%	%\vspace{-0.35in}
%	\caption{Video surveillance traffic capture scenario.}
%	\label{fig:test_traffic}
%\end{figure}

\begin{table}[t]
	\renewcommand{\arraystretch}{1.3}
	\caption{Amount of packets per burst depending on camera compression.}
	\label{table:camera_packets}
	\centering
	\begin{tabular}{lcc}
		\hline
		\hline
		$ Resolucition $ & $ Compression \, level $ & $ Packets \, per \, burst $\\
		\hline 
		 & $ 50 \; Kbytes $ & $ 41 $ \\ 
		\rowcolor[gray]{0.9}$ 704\times576 \; pixeles $ & $ 32 \; Kbytes $ & $ 26 $ \\
		 & $ 16 \; Kbytes $ & $ 10 $ \\ 
		 \hline
		\multirow{2}{2.5cm}{$ 352\times288 \; pixeles $} & $ 13 \; Kbytes $ & $ 9 $ \\ 
		%\cline{2-3}
		 & $ 4 \; Kbytes $ & $ 3 $ \\ 
		\hline 
		\hline
	\end{tabular} 
\end{table}

Voice traffic was generated according to $ G.729 $ recommendation ($ 20 \, ms $ for inter-packet time and $ 2 $ samples per packet), resulting on a packet size of $ 60 \; bytes $.

As in a real scenario, flows do not start at the same time, there is a starting period in which all flows starts randomly. So, bursts have not been forced, they appear by the specific application behaviour or by the overlapping in the combination of the different applications traffic. Packet loss differs for each test because in some cases increases the overlapping flows. We have obtained accuracy average results when each test was repeated $ 40 $ times. Simulation time is $ 60 \; s $ for each tests.

\subsection{First scenario}

The first scenario is shown in figure \ref{fig:scenarios}. In this case two and three different camera communications share the same Internet access limited to $ 3.5 \; Mbps $. This bandwidth has been chosen in order to set the offered bandwidth to $ 85\% $ of the link capacity when three cameras are present. The main aim of the test is to determinate the packet loss rate in the mixed bursty traffic for different buffer sizes.

%\begin{figure}[ht]
%	\centering
%	\includegraphics[height=2in,width=0.45\textwidth]{Images/test_camera.jpg}
%	%\vspace{-0.35in}
%	\caption{First scenario: two and three camera connections.}
%	\label{fig:cameras}
%\end{figure}

\begin{figure}[t]
	\centering
	\includegraphics[height=3in,width=0.45\textwidth]{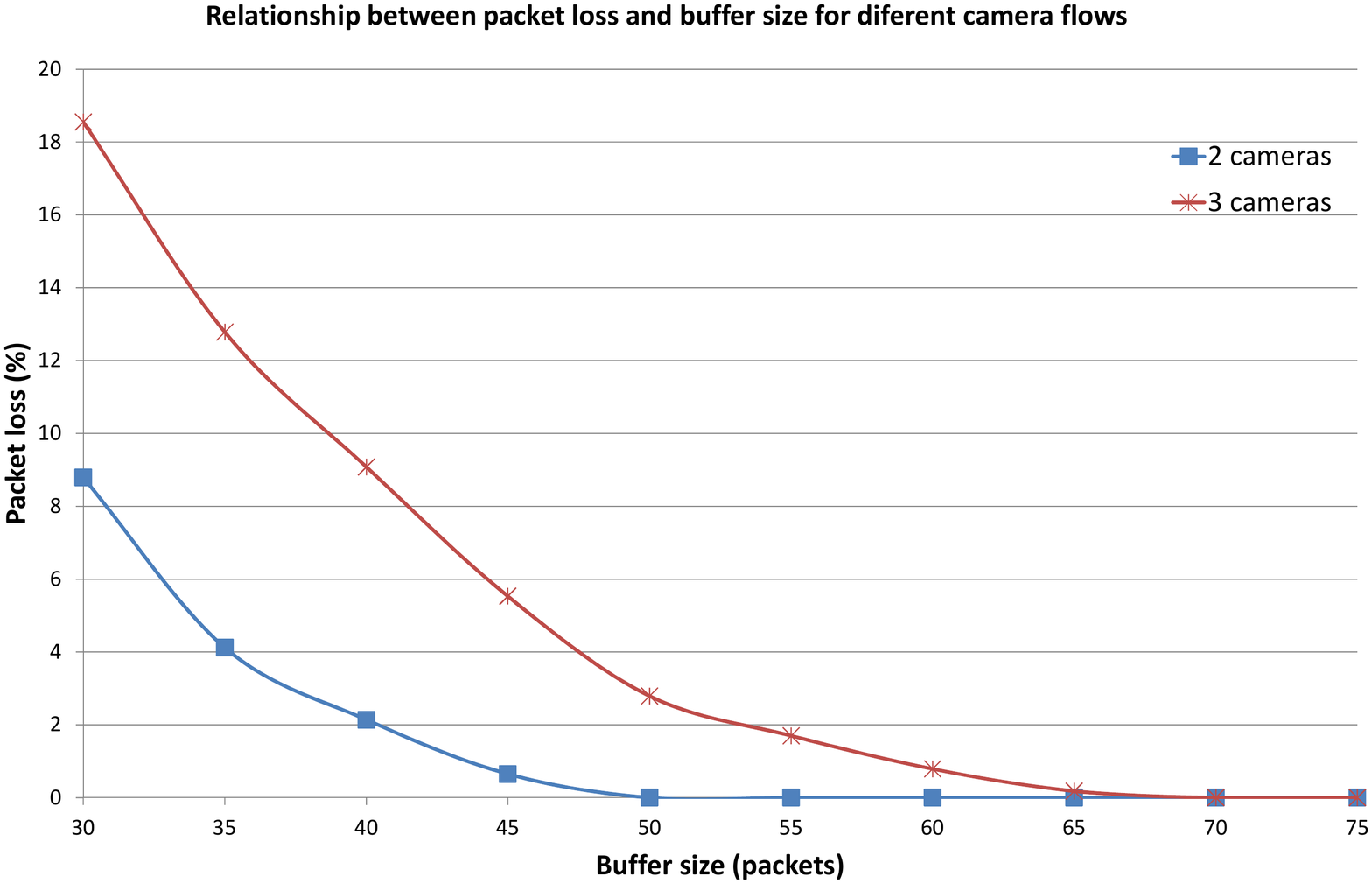}
	\caption{Packet loss for two and three camera flows for different buffer size.}
	\label{fig:camera}
\end{figure}

Although the offered traffic is roughly is about $ 57\% $ and $ 85\% $ for two and three cameras respectively, packet loss may be unacceptable, as shown in figure \ref{fig:camera}; the cause is that the amount of packets per burst exceeds the capacity of the buffer when more than one connection is set. As an example, we can think about a buffer size of $ 30 $ packets. If the number of packets in a burst is $ 26 $, it would be easy that the buffer gets full whenever a burst arrives. Furthermore, the relationship between packet loss and the number of camera flows is not linear. This relationship decreases when buffer size increases.

%At the same time, other interesting phenomenon can be observed: when comparing the $ 10 $ and $ 100 \; Mbps $ results of figure \ref{fig:camera}, we see that packet loss is higher when network speed is $ 100 \; Mbps $. Furthermore, there are some cases (e.g, $ 2 $ cameras and buffer of $ 45 $ packets) in which packet loss only appears for the fastest network. Why is this happening? If the capacity of the Internet connection remains the same, a $ 100 \; Mbps $ network will fill the buffer faster than a $ 10 \; Mbps $ one whenever a burst is generated by a camera. In these cases, upgrading the network speed from $ 10 $ to $ 100 \; Mbps $ may worsen the network performance.

\begin{figure*}[ht]
	\centering
	\subfigure[Packet loss by flows.]{\includegraphics[height=3in,width=0.45\textwidth]{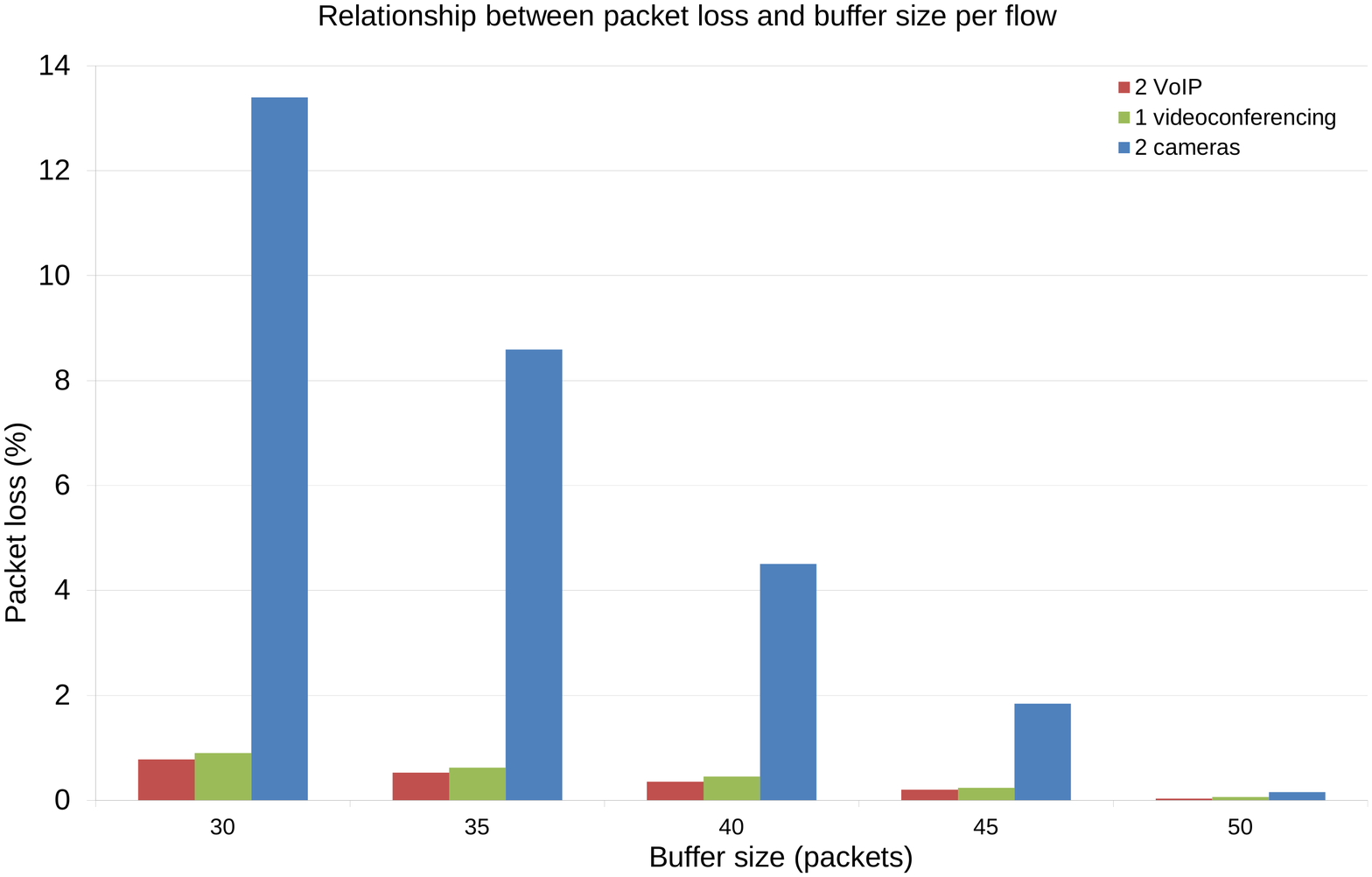}}
	\subfigure[Packet loss distribution by flows.]{\includegraphics[height=3in,width=0.45\textwidth]{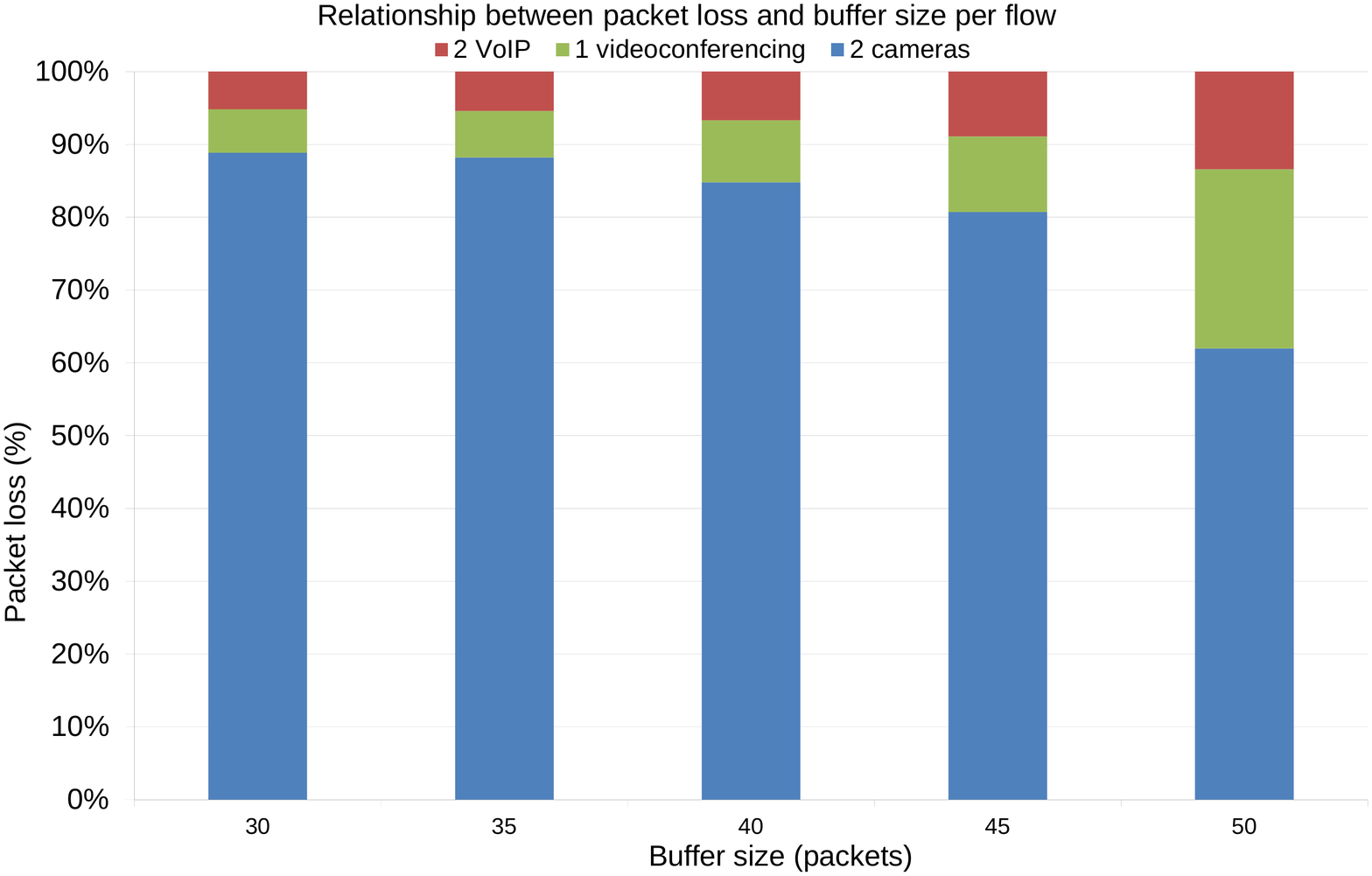}}
	%\vspace{-0.2in}
	\caption{Packet loss when link utilization is $ 70\% $ for different buffer size.}
	\label{fig:scenario_70}
\end{figure*}

\begin{figure*}[ht]
	\centering
	\subfigure[Packet loss by flows.]{\includegraphics[height=3in,width=0.45\textwidth]{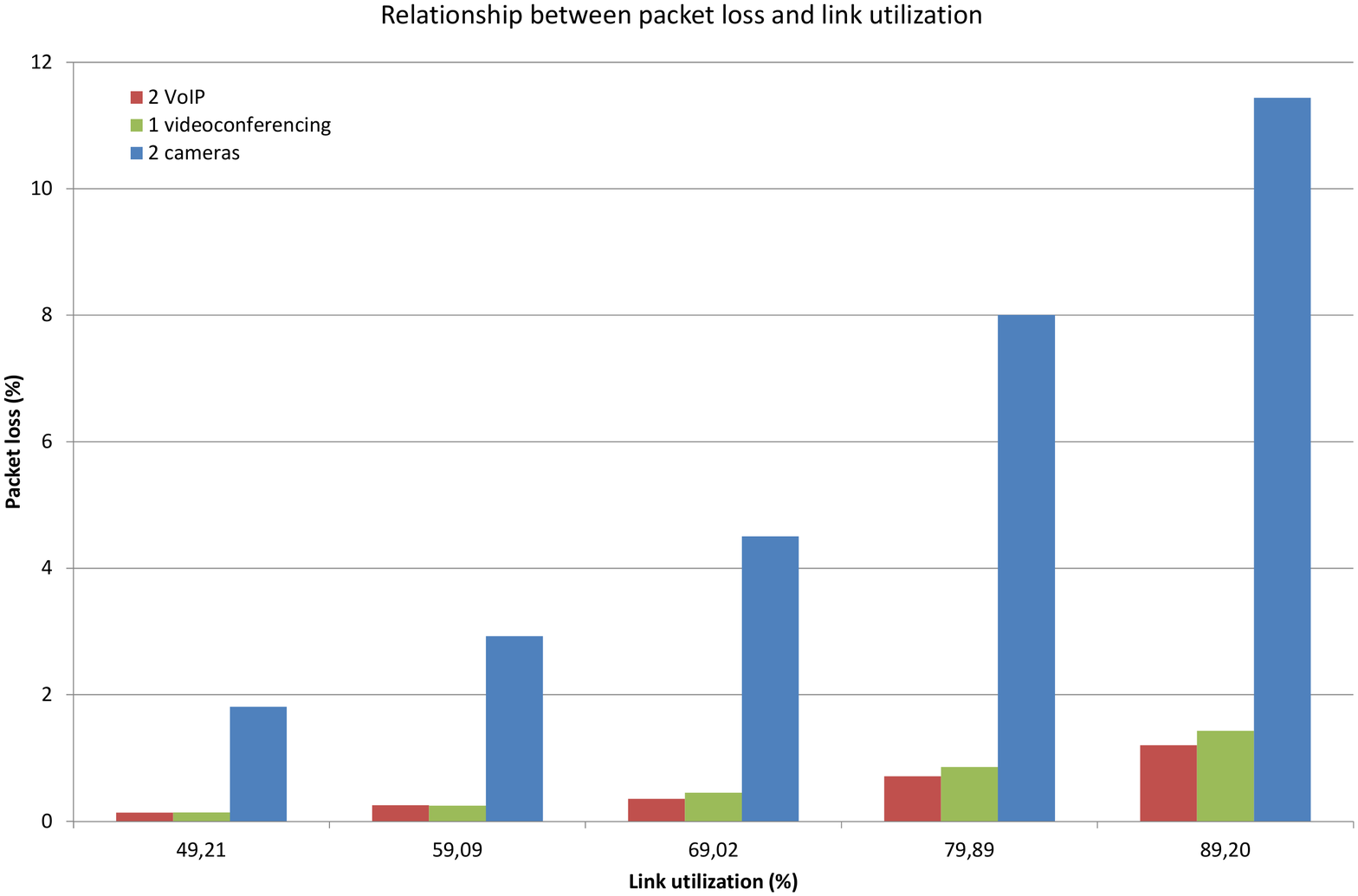}}
	\subfigure[Packet loss distribution by flows.]{\includegraphics[height=3in,width=0.45\textwidth]{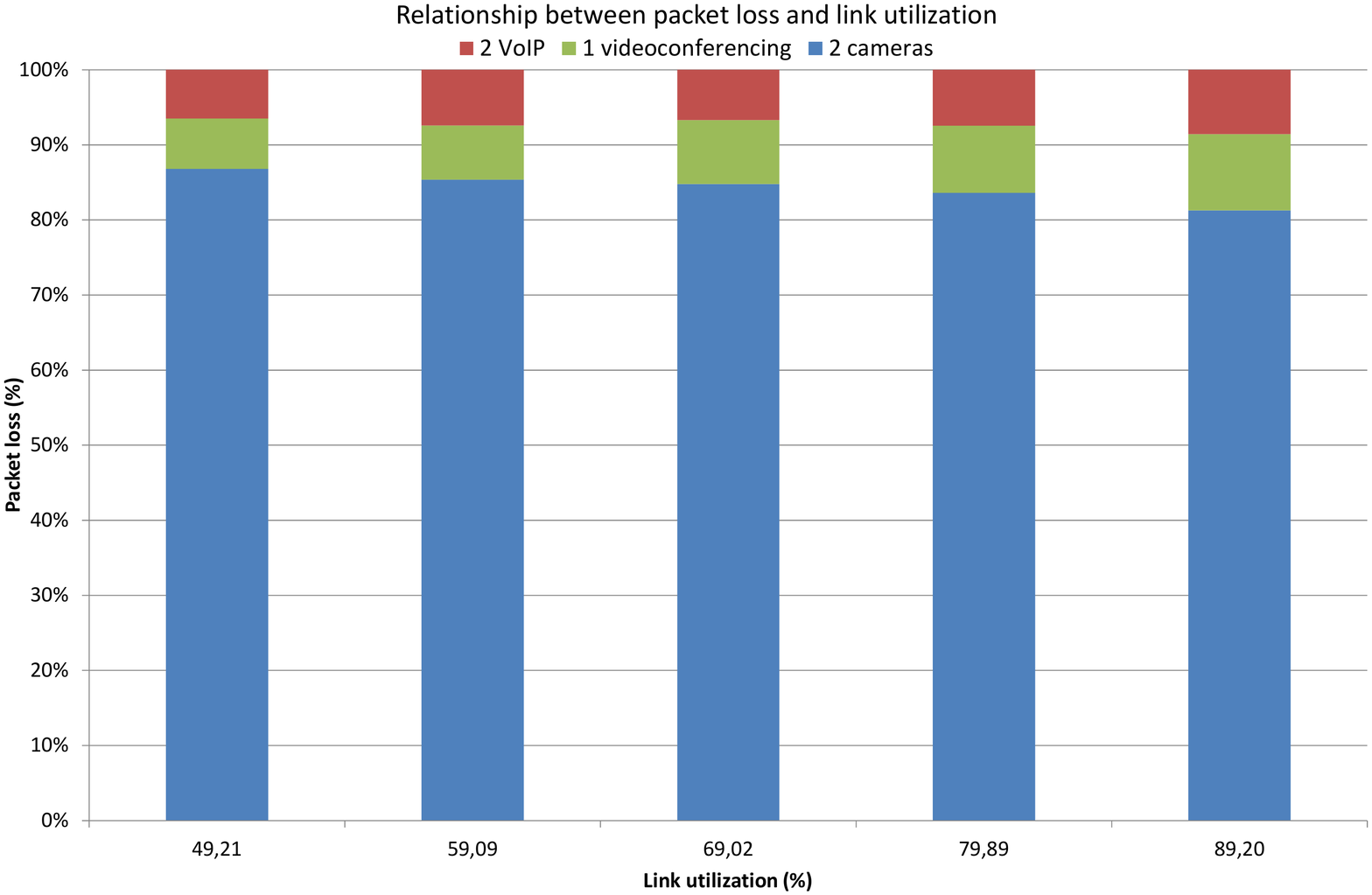}}
	%\vspace{-0.2in}
	\caption{Packet loss when buffer size is $ 40 $ packets for different values of link utilization.}
	\label{fig:scenario_40}
\end{figure*}

\subsection{Second scenario}

The second scenario is shown in figure \ref{fig:scenarios}. Two IP camera flows, one videoconferencing session and two VoIP calls are used as test traffic, so the total bandwidth generated is $ 3.5 \; Mbps $. 

\subsubsection{General study}

Two different tests have been performed in this scenario: in the first one, the Internet access link has been set to $ 5 \; Mbps $, so average link utilization is fixed ($ 70\% $) and different values of the buffer size are tested. In the second tests, the buffer size of the Internet access router is fixed at $ 40 $ packets and the simulations are run using different values of the access bandwidth and consequently different levels of link utilization, ranging from $ 50\% $ to $ 90\% $. The presented results are for the aggregate traffic of the three applications. 

%\begin{figure}[ht]
%	\centering
%	\includegraphics[height=2.5in,width=0.45\textwidth]{Images/test_40_70_v2.jpg}
%	%\vspace{-0.35in}
%	\caption{Second scenario: two camera connections, videoconferencing and two VoIP calls.}
%	\label{fig:scenario}
%\end{figure}

For the first case, the packet loss per flow can be observed in figure \ref{fig:scenario_70}. Packet loss affects all the applications, so the presence of a bursty application (video surveillance) causes packet loss for all the coexisting applications, even for those generating constant bit rate traffic (VoIP). In addition, packet loss decreases when buffer size is increased, because big buffers can absorb the burst produced by the traffic mix. On the other hand, packet loss distribution is not the same for all buffers tested, although there is a small packet loss for big buffers, the percent of losses increases for videoconferencing and VoIP.

As expected, packet loss increases when link utilization grows in the case of $ 40 $ packets buffer (figure \ref{fig:scenario_40}). Again, packet loss distribution is not the same and the percent of losses increases for videoconferencing and VoIP.

\subsubsection{Deep study}

\begin{figure*}[ht]
	\centering
	\includegraphics[height=3in,width=1\textwidth]{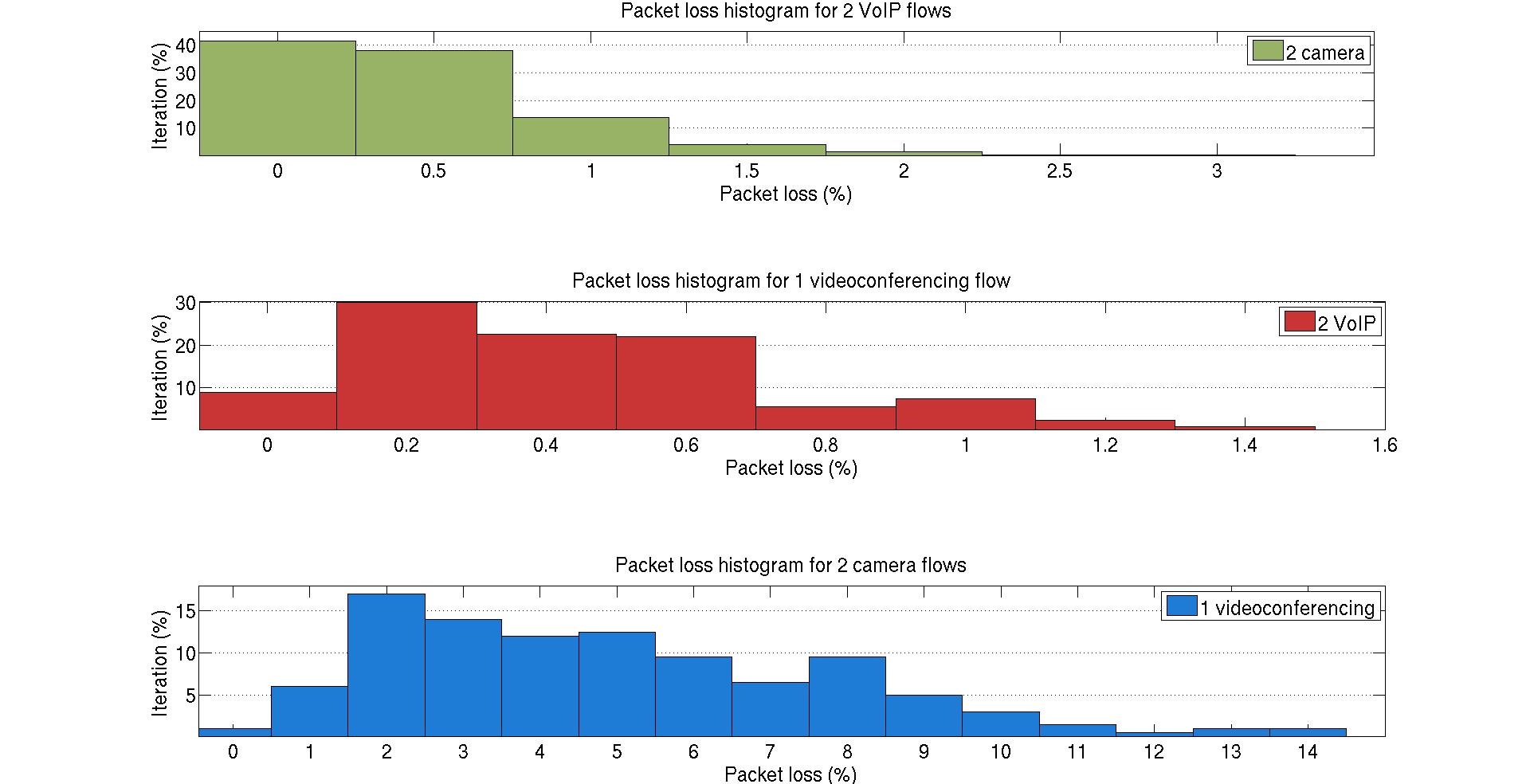}
	\caption{Packet loss histogram for different multimedia flows with a buffer size of $ 40 $ packets and $ 70\% $ link utilization}
	\label{fig:histogram}
\end{figure*}

In order to deeply analyze the effect of packet loss for each flow, we have selected a scenario with a $ 70\% $ link utilization and a buffer size of $ 40 $ packets and the same mentioned applications. In this specific scenario, tests were repeated $ 200 $ times to observe the effect of the overlapping flows cited above and their relationship with packet loss. The results are presented by a histogram in figure \ref{fig:histogram} in which the y axis represents the amount of iterations and x axis shows the packet loss ranges.

Packet loss presents different values for different iterations because in some cases the flow overlapping is bigger. This phenomenon mainly harms sensitive traffic as VoIP. Almost $ 80\% $ of the calls present a packet loss smaller than $ 0.75\% $. Packet loss reaches $ 3\% $ in $ 0.5\% $ of the cases (equivalent to $ 20 $ calls).
%
%\begin{itemize}
%\item \textbf{Como ejemplo hemos obtenido el MOS para las 200 iteraciones y el histograma se muestra en la figura \ref{fig:mos}}
%\item \textbf{El histograma muestra que cierta cantidad de llamadas han tenido un MOS regular apesar de tener suficiente ancho de banda}
%\item \textbf{La probabilidad de que una llamada tenga un MOS o mayor se muestra en las figuras \ref{fig:mos_0-40_1} y \ref{fig:mos_0-40_2}, donde se observa que dicha probabilidad disminuye con el aumento del retardo de la red}
%\end{itemize}

\begin{figure*}[ht]
	\centering
	\includegraphics[height=2.5in,width=\textwidth]{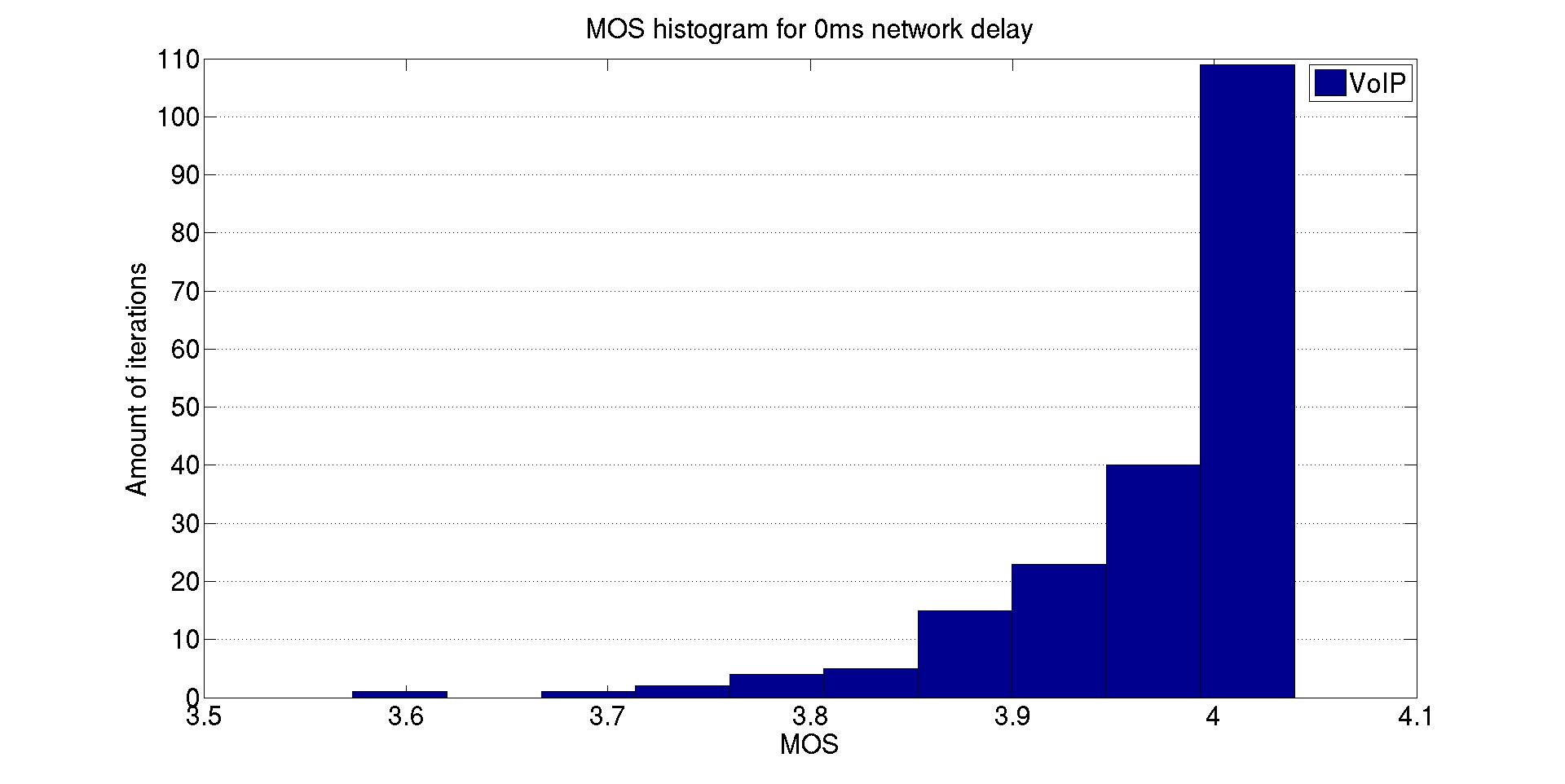}
	\caption{MOS histogram for 2 VoIP communications with a buffer size of $ 40 $ packets and $ 70\% $ link utilization.}
	\label{fig:mos}
\end{figure*}

On the other hand, this tool provides a means to estimate the subjective Mean Opinion Score (MOS) rating of voice quality over these planned network environments. For this reason, we have obtained the MOS for each iteration with the aim of comparing the quality in each case. The results are presented by a histogram. The chart in figure \ref{fig:mos} shows a significant amount of calls with a \textit{medium quality} according to \cite{mos} in a scenario with optimal conditions to obtain the \textit{best quality}.

\begin{figure*}[ht]
	\centering
	\includegraphics[height=3in,width=\textwidth]{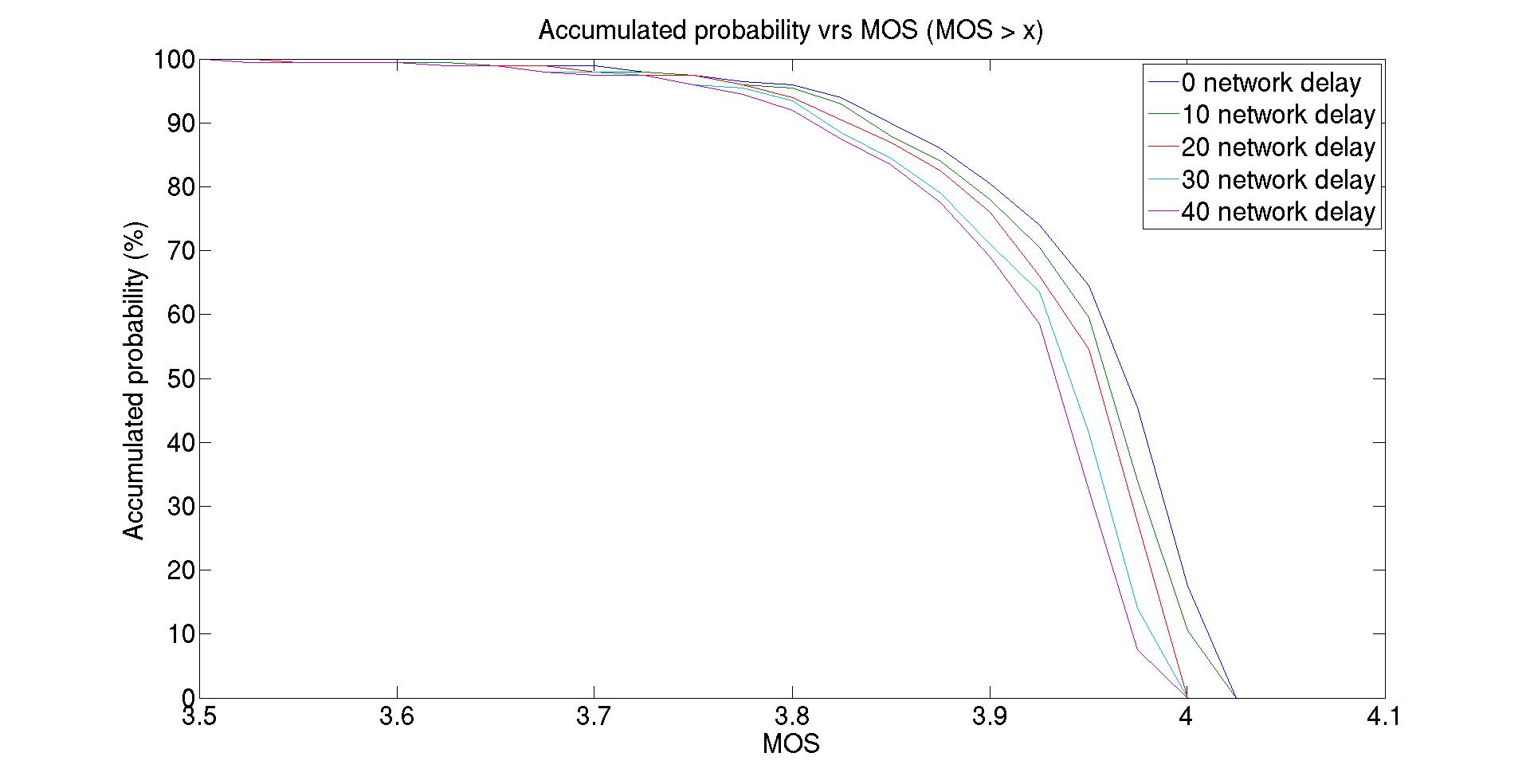}
	\caption{Accumulative MOS probability for different network delay with a buffer size of $ 40 $ packets and $ 70\% $ link utilization.}
	\label{fig:mos_0-40_1}
\end{figure*}

Additional useful information can be obtained from the accumulative MOS probability (see figure \ref{fig:mos_0-40_1}) in which the MOS has been calculated for $ 5 $ different values of the network delay. The results show that about $ 98\% $ of the calls can obtain a MOS of $ 3.6 $ or can never, it is equivalent to \textit{medium quality}. A \textit{high quality} can never be reached even if the network delay is $ 0ms $.

Accumulative MOS probability decreases quickly in \textit{medium quality} ($ 3.60-4.03 $) and cannot reach a higher quality. Figure \ref{fig:mos_0-40_1} illustrates this phenomenon.

\section{Conclusion}

This paper has studied packet loss caused by the router buffer, in the presence of applications generating bursty traffic, and their influence on VoIP subjective quality. Tests in two scenarios using NS-2 with real traces of different multimedia applications were presented, always with medium link utilization. 

The buffer size has been identified as a critical parameter for network planning in these environments. The reason is that the relationship between buffer size and burst length has to be coherent in order to allocate all the packets thus avoiding packet discarding because the numbers of packets in the burst exceeds the capacity of the buffer.

In addition, it has been observed that bursty traffic affects the other applications sharing the same link. In order to show the effect of the bursty nature of these applications, we have measured the MOS of concurrent VoIP calls. The results show that VoIP calls are only able to obtain a \textit{medium quality}, failing to reach better qualities even when link utilization is set to $ 70\% $.


\section*{Acknowledgment}

This work has been partially financed by the European Social Fund in collaboration with the Government of Arag\'{o}n, CPUFLIPI Project (MICINN TIN2010-17298), Ibercaja Obra Social, Project of C\'atedra Telef\'onica, University of Zaragoza, Banco Santander and Fundaci\'on Carolina.



\bibliographystyle{IEEEtran}
\bibliography{IEEEabrv,citas}

\begin{thebibliography}{10}
\providecommand{\url}[1]{#1}
\csname url@samestyle\endcsname
\providecommand{\newblock}{\relax}
\providecommand{\bibinfo}[2]{#2}
\providecommand{\BIBentrySTDinterwordspacing}{\spaceskip=0pt\relax}
\providecommand{\BIBentryALTinterwordstretchfactor}{4}
\providecommand{\BIBentryALTinterwordspacing}{\spaceskip=\fontdimen2\font plus
\BIBentryALTinterwordstretchfactor\fontdimen3\font minus
  \fontdimen4\font\relax}
\providecommand{\BIBforeignlanguage}[2]{{%
\expandafter\ifx\csname l@#1\endcsname\relax
\typeout{** WARNING: IEEEtran.bst: No hyphenation pattern has been}%
\typeout{** loaded for the language `#1'. Using the pattern for}%
\typeout{** the default language instead.}%
\else
\language=\csname l@#1\endcsname
\fi
#2}}
\providecommand{\BIBdecl}{\relax}
\BIBdecl

\bibitem{games3}
G.~Huang, M.~Ye, and L.~Cheng, ``Modeling system performance in mmorpg,'' in
  \emph{Global Telecommunications Conference Workshops, 2004. GlobeCom
  Workshops 2004. IEEE}, nov.-3 dec. 2004, pp. 512 -- 518.

\bibitem{camera1}
S.~Fleck and W.~Strasser, ``Smart camera based monitoring system and its
  application to assisted living,'' \emph{Proceedings of the IEEE}, vol.~96,
  no.~10, pp. 1698 --1714, oct. 2008.

\bibitem{buffers5}
\BIBentryALTinterwordspacing
A.~Vishwanath, V.~Sivaraman, and M.~Thottan, ``Perspectives on router buffer
  sizing: recent results and open problems,'' \emph{SIGCOMM Comput. Commun.
  Rev.}, vol.~39, pp. 34--39, March 2009. [Online]. Available:
  \url{http://doi.acm.org/10.1145/1517480.1517487}
\BIBentrySTDinterwordspacing

\bibitem{buffers2}
\BIBentryALTinterwordspacing
A.~Vishwanath, V.~Sivaraman, and G.~N. Rouskas, ``Considerations for sizing
  buffers in optical packet switched networks,'' \emph{IEEE INFOCOM 2009 The
  28th Conference on Computer Communications}, pp. 1323--1331, 2009. [Online].
  Available:
  \url{http://ieeexplore.ieee.org/lpdocs/epic03/wrapper.htm?arnumber=5062047}
\BIBentrySTDinterwordspacing

\bibitem{buffers6}
\BIBentryALTinterwordspacing
C.~Villamizar and C.~Song, ``High performance tcp in ansnet,'' \emph{SIGCOMM
  Comput. Commun. Rev.}, vol.~24, pp. 45--60, October 1994. [Online].
  Available: \url{http://doi.acm.org/10.1145/205511.205520}
\BIBentrySTDinterwordspacing

\bibitem{buffers7}
\BIBentryALTinterwordspacing
G.~Appenzeller, I.~Keslassy, and N.~McKeown, ``Sizing router buffers,''
  \emph{SIGCOMM Comput. Commun. Rev.}, vol.~34, pp. 281--292, August 2004.
  [Online]. Available: \url{http://doi.acm.org/10.1145/1030194.1015499}
\BIBentrySTDinterwordspacing

\bibitem{buffers8}
\BIBentryALTinterwordspacing
N.~Beheshti, Y.~Ganjali, M.~Ghobadi, N.~McKeown, and G.~Salmon, ``Experimental
  study of router buffer sizing,'' pp. 197--210, 2008. [Online]. Available:
  \url{http://doi.acm.org/10.1145/1452520.1452545}
\BIBentrySTDinterwordspacing

\bibitem{buffers10}
\BIBentryALTinterwordspacing
M.~Enachescu, Y.~Ganjali, A.~Goel, N.~McKeown, and T.~Roughgarden, ``Part iii:
  routers with very small buffers,'' \emph{SIGCOMM Comput. Commun. Rev.},
  vol.~35, pp. 83--90, July 2005. [Online]. Available:
  \url{http://doi.acm.org/10.1145/1070873.1070886}
\BIBentrySTDinterwordspacing

\bibitem{buffers9}
\BIBentryALTinterwordspacing
A.~Vishwanath and V.~Sivaraman, ``{Routers With Very Small Buffers: Anomalous
  Loss Performance for Mixed Real-Time and TCP Traffic},'' pp. 80--89, Jun.
  2008. [Online]. Available: \url{http://dx.doi.org/10.1109/IWQOS.2008.16}
\BIBentrySTDinterwordspacing

\bibitem{buffers1}
\BIBentryALTinterwordspacing
J.~Sommers, P.~Barford, A.~Greenberg, and W.~Willinger, ``An sla perspective on
  the router buffer sizing problem,'' \emph{SIGMETRICS Perform. Eval. Rev.},
  vol.~35, pp. 40--51, March 2008. [Online]. Available:
  \url{http://doi.acm.org/10.1145/1364644.1364645}
\BIBentrySTDinterwordspacing

\bibitem{yo1}
L.~Sequeira, J.~Fernandez-Navajas, J.~Saldana, and L.~Casadesus, ``Empirically
  characterizing the buffer behaviour of real devices,'' in \emph{Performance
  Evaluation of Computer and Telecommunication Systems (SPECTS), 2012
  International Symposium on}, july 2012, pp. 1 --6.

\bibitem{bursty1}
H.~Jiang and C.~Dovrolis, ``Why is the internet traffic bursty in short time
  scales?'' in \emph{ACM SIGMETRICS Performance Evaluation Review}, vol.~33,
  no.~1.\hskip 1em plus 0.5em minus 0.4em\relax ACM, 2005, pp. 241--252.

\bibitem{gtc17}
\BIBentryALTinterwordspacing
J.~Saldana, J.~Fern\'{a}ndez-Navajas, J.~Ruiz-Mas, J.~Murillo, E.~V. Navarro,
  and J.~I. Aznar, ``Evaluating the influence of multiplexing schemes and
  buffer implementation on perceived voip conversation quality,''
  \emph{Computer Networks}, vol.~56, no.~7, pp. 1893 -- 1919, 2012. [Online].
  Available:
  \url{http://www.sciencedirect.com/science/article/pii/S1389128612000618}
\BIBentrySTDinterwordspacing

\bibitem{emodel}
ITU-T, ``The e-model: a computational model for use in transmission planning,''
  Tech. Rep., Dicember 2011.

\bibitem{mos}
\BIBentryALTinterwordspacing
R.~G. Cole and J.~H. Rosenbluth, ``Voice over ip performance monitoring,''
  \emph{SIGCOMM Comput. Commun. Rev.}, vol.~31, no.~2, pp. 9--24, Apr. 2001.
  [Online]. Available: \url{http://doi.acm.org/10.1145/505666.505669}
\BIBentrySTDinterwordspacing

\bibitem{games5}
\BIBentryALTinterwordspacing
A.~F. Wattimena, R.~E. Kooij, J.~M. van Vugt, and O.~K. Ahmed, ``Predicting the
  perceived quality of a first person shooter: the quake iv g-model,'' in
  \emph{Proceedings of 5th ACM SIGCOMM workshop on Network and system support
  for games}, ser. NetGames '06.\hskip 1em plus 0.5em minus 0.4em\relax New
  York, NY, USA: ACM, 2006. [Online]. Available:
  \url{http://doi.acm.org/10.1145/1230040.1230052}
\BIBentrySTDinterwordspacing

\bibitem{gtc14}
J.~Saldana, J.~Fern\'{a}ndez-Navajas, J.~Ruiz-Mas, E.~Viruete~Navarro, and
  L.~Casadesus, ``Influence of online games traffic multiplexing and router
  buffer on subjective quality,'' \emph{in Proc. CCNC 2012- 4th IEEE
  International Workshop on Digital Entertainment, Networked Virtual
  Environments, and Creative Technology (DENVECT)}, pp. 482--486, Las Vegas,
  January 2012.

\bibitem{p2p4}
J.~M. Saldana, J.~Fernandez-Navajas, J.~Ruiz-Mas, E.~V. Navarro, and
  L.~Casadesus, ``The utility of characterizing packet loss as a function of
  packet size in commercial routers,'' in \emph{CCNC}, 2012, pp. 346--347.

\bibitem{games4}
W.-C. Feng, F.~Chang, W.-C. Feng, and J.~Walpole, ``Provisioning on-line games:
  a traffic analysis of a busy counter-strike server,'' in \emph{Proceedings of
  the 2nd ACM SIGCOMM Workshop on Internet measurment}.\hskip 1em plus 0.5em
  minus 0.4em\relax ACM, 2002, pp. 151--156.

\bibitem{p2p3y18}
B.~Fallica, Y.~Lu, F.~Kuipers, R.~Kooij, and P.~V. Mieghem, ``On the quality of
  experience of sopcast,'' in \emph{Next Generation Mobile Applications,
  Services and Technologies, 2008. NGMAST'08. The Second International
  Conference on}, September 2008, pp. 501--506.

\end{thebibliography}
%\begin{thebibliography}{1}
%\end{thebibliography}

\end{document}